\documentclass[8.5pt,twoside,twocolumn]{article}
\usepackage[super,sort&compress,comma]{natbib}
\usepackage{mhchem}
\usepackage{times,mathptmx}
\usepackage{sectsty}
\usepackage{balance}

\usepackage{graphicx} 
\usepackage{lastpage}
\usepackage[format=plain,justification=raggedright,singlelinecheck=false,font=small,labelfont=bf,labelsep=space]{caption}
\usepackage{fancyhdr}

\usepackage{subfig}
\usepackage{SIunits}
\begin{document}

\thispagestyle{plain}
\fancypagestyle{plain}{
\renewcommand{\headrulewidth}{1pt}}
\renewcommand{\thefootnote}{\fnsymbol{footnote}}
\renewcommand\footnoterule{\vspace*{1pt}%
\hrule width 3.4in height 0.4pt \vspace*{5pt}}
\setcounter{secnumdepth}{5}

\makeatletter
\def\subsubsection{\@startsection{subsubsection}{3}{10pt}{-1.25ex plus -1ex minus -.1ex}{0ex plus 0ex}{\normalsize\bf}}
\def\paragraph{\@startsection{paragraph}{4}{10pt}{-1.25ex plus -1ex minus -.1ex}{0ex plus 0ex}{\normalsize\textit}}
\renewcommand\@biblabel[1]{#1}
\renewcommand\@makefntext[1]%
{\noindent\makebox[0pt][r]{\@thefnmark\,}#1}
\makeatother
\renewcommand{\figurename}{\small{Fig.}~}
\sectionfont{\large}
\subsectionfont{\normalsize}

\fancyfoot{}
\fancyfoot[RO]{\footnotesize{\sffamily{1--\pageref{LastPage} ~\textbar  \hspace{2pt}\thepage}}}
\fancyfoot[LE]{\footnotesize{\sffamily{\thepage~\textbar\hspace{3.45cm} 1--\pageref{LastPage}}}}
\fancyhead{}
\renewcommand{\headrulewidth}{1pt}
\renewcommand{\footrulewidth}{1pt}
\setlength{\arrayrulewidth}{1pt}
\setlength{\columnsep}{6.5mm}
\setlength\bibsep{1pt}

\twocolumn[
  \begin{@twocolumnfalse}
\noindent\LARGE{\textbf{Tunable light trapping and absorption enhancement with graphene ring arrays}}
\vspace{0.6cm}

\noindent\large{\textbf{Shuyuan Xiao,\textit{$^{a}$} Tao Wang,$^{\ast}$ \textit{$^{a}$} Yuebo Liu,\textit{$^{b}$} Chen Xu,\textit{$^{c}$} Xu Han,\textit{$^{a}$} and Xicheng Yan\textit{$^{a}$}}}\vspace{0.5cm}

\noindent\textit{\small{\textbf{Received Xth XXXXXXXXXX 20XX, Accepted Xth XXXXXXXXX 20XX\newline
First published on the web Xth XXXXXXXXXX 200X}}}

\noindent \textbf{\small{DOI: 10.1039/b000000x}}
\vspace{0.6cm}

\noindent \normalsize{Surface plasmon resonance (SPR) has been intensively studied and widely employed for light trapping and absorption enhancement. In the mid-infrared and terahertz (THz) regime, graphene supports the tunable SPR via manipulating its Fermi energy and enhances light-matter interaction at the selective wavelength. In this work, periodic arrays of graphene rings are proposed to introduce tunable light trapping with good angle polarization tolerance and enhance the absorption in the light-absorbing materials nearby to more than one order. Moreover, the design principle here could be set as a template to achieve multi-band plasmonic absorption enhancement by introducing more graphene concentric rings into each unit cell. This work not only opens up new ways of employing graphene SPR, but also leads to practical applications in high-performance simultaneous multi-color photodetection with high efficiency and tunable spectral selectivity.}
\vspace{0.5cm}
 \end{@twocolumnfalse}
  ]

\section{Introduction}\label{sec1}


\footnotetext{\textit{$^{a}$~Wuhan National Laboratory for Optoelectronics, Huazhong University of Science and Technology, Wuhan 430074, People's Republic of China. E-mail: wangtao@hust.edu.cn}}
\footnotetext{\textit{$^{b}$~School of Information and Optoelectronic Science and Engineering, South China Normal University, Guangzhou 510006, People's Republic of China}}
\footnotetext{\textit{$^{c}$~Department of Physics, New Mexico State University, Las Cruces 88001, United State of America}}


In recent years, the light absorption has demonstrated a variety of optoelectronic applications in photovoltaic cells\cite{kelzenberg2010enhanced,huang2013enhancing,dessi2015thiazolo}, spatial light modulation \cite{liu2010infraredspatial,savo2014liquid}, ultrasensitive biosensing\cite{chen2012dual,wang2015novel,chen2016catalase,lin2016novel}, photocatalytic reaction\cite{chen2011increasing,abdi2014plasmonic,zhou2015plasmon,wang2015hybrid} and photodetection\cite{knight2011photodetection,li2012facile,sharma2015detection}. The extinction of the light-absorbing semiconductors employed in these devices is expected to be high enough to guarantee sufficient light absorption. On the other hand, there is also urgent need to reduce the thickness and volume of them to decrease the resource consumption, reduce the material deposition requirement, and improve the integrability, which also inevitably weaken the interaction between light and the semiconductors\cite{callahan2012solar,liu2015perfect}. Moreover, the emergency of two-dimensional (2D) materials with unique properties has stimulated great interests in scientific studies and technological developments \cite{bonaccorso2010graphene,nulakani2015coro,mak2010atomically,wang2015physical,zhang2016strong,fang2013tuning}, but the innate atomic thickness greatly hampers the utilization of light \cite{xia2014two,li2016hybrid}. As a result, enhancing the light absorption in the above-mentioned semiconductors and 2D materials to satisfy the requirements of practical applications, especially of some light-driven-related applications such as photodetection, has become of increasing importance. Surface plasmonic resonance (SPR), the collective electronic excitation at the interface between a metal and a dielectric medium, has been introduced as one of the most widely used approaches to enhance the light absorption for it can strongly couple to the incidence light and lead to efficient light trapping and absorption enhancement in the subwavelength scale structure\cite{schaadt2005enhanced,liu2010infrared,choi2013versatile,bohn2014design}. Based on this, a range of plasmonic structures such as metal strip\cite{le2009plasmon,song2013great,cai2015tunable}, disk\cite{heeg2012polarized,wang2013light}, ring\cite{khardikov2010trapping,lu2015highly}, cross\cite{vasic2013graphene} and other shapes\cite{dang2013tunable,xiong2015ultrabroadband} have been proposed to integrate with the light-absorbing materials to improve their absorption performances. However, the spectral responses of these metal SPRs are dominated by the geometric parameters, and the operation wavelength will be unchangeable once these devices are manufactured, which still restricts flexible applications in practice.

Graphene is a monolayer carbon material that has become a promising building block in the state of the art technology due to its remarkable mechanical, thermal and electromagnetic properties\cite{neto2009electronic,geim2009graphene,novoselov2012roadmap,chung2016electronic}. As a typical example of 2D materials, the interaction between light and graphene is quite weak and a single sheet of graphene can absorb only $2.3\%$ energy of the incidence light in the visible and near infrared. Unlike other 2D materials, graphene itself behaves like metals when coupling with the incidence light and supports SPR in the mid-infrared and terahertz (THz) region, therefore benefits to light trapping and absorption enhancement\cite{grigorenko2012graphene,brar2013highly,jang2014tunable}. Moreover, compared with metal structure, the continuously tunable surface conductivity of graphene via manipulating its Fermi energy by electric gating or chemical doping enable dynamically tunable resonance\cite{he2015tunable,he2015investigation,lin2015combined,han2015dynamically,liu2016actively}, which can be used to amplify the photoresponse to the incidence light at the selective wavelength, making it possible for highly accurate photodetection\cite{fang2012graphene}. Tunable single-\cite{nikitin2012surface}, dual-\cite{ke2015plasmonic} and multi-band\cite{xu2013novel} absorption enhancement with graphene plasmonics have been realized and even perfect absorption also achieved with graphene absorber backed by a gold mirror\cite{liu2013dual,zhang2014grapheneabsorber,yao2016dual}. However, most of these previous works concentrated on the absorption enhancement in graphene itself while little attention has been paid to trap light and enhance the light absorption in other light-absorbing materials with graphene SPR.

To demonstrate this kind of use of graphene SPR, a simulation study on tunable light trapping and absorption enhancement has been systematically conducted in this work with a hybrid periodic array composed of a graphene ring on the top of the light-absorbing materials (thin film semiconductor or 2D material) separated by an insulating layer. Graphene ring shows good operation angle polarization tolerance due to its high symmetry, and it possesses the advantage of more free geometric parameters compared with disk\cite{liu2012tunable,li2014sensitive}. The simulation results show that the excitations of SPR in the graphene ring trap a sizable part of the incidence light and significantly enhance the light absorption in surrounding absorbing materials. With manipulating the Fermi energy of graphene and adjusting the geometric parameters, the absorption could be modulated over a large range. In addition, our proposed structure leaves room for introducing more graphene rings into the unit cell to form multiple resonating structure, and consequently enables dynamically tunable multi-band absorption enhancement in the below light-absorbing materials, which is considered to play a significant role in the simultaneous multi-color photodetection with high efficiency and tunable spectral selectivity.

\section{The geometric structure and numerical model}\label{sec2}
The schematic geometry of our proposed structure is shown in Figure 1. The unit cell is arranged in a periodic array with a lattice constant $P=300$ nm and composed of a graphene ring on the top of the absorbing layer separated by an insulating layer. $R=75$ nm is the radius of the graphene ring with a width $W=30$ nm and the effective thickness is set as $t_g=1$ nm. The thicknesses of the insulating layer and the absorbing layer are $t_i=20$ nm and $t_a=100$ nm, and the substrate is assumed to be semi-infinite. Both the insulating layer and the substrate are considered as lossless dielectric with a real permittivity of $\varepsilon_d=1.96$. Comparable with some typical materials employed for photodetection in the mid-infrared and THz regime such as HgCdTe (MCT)\cite{palik1991handbook}, the absorbing layer is modeled as thin film semiconductor through a complex permitivity of $\varepsilon_a=\varepsilon^{'}+i\varepsilon^{''}$, where $\varepsilon^{'}=10.9$ and $\varepsilon^{''}$ is related to the attenuation coefficient $\alpha=-(2\pi/\lambda)\text{Im}(\sqrt{\varepsilon^{'}+i\varepsilon^{''}})$ accounting for the loss.
\begin{figure}[htbp]\label{fig:1}
\centering
\includegraphics[scale=0.4]{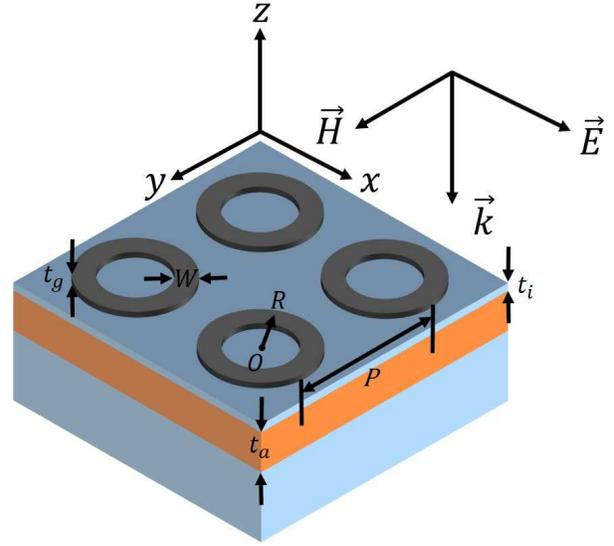}
\caption{The schematic geometry of our proposed hybrid periodic array. Each unit cell is composed of a graphene ring on the top of the absorbing layer separated by an insulating layer.}
\end{figure}

Graphene is considered as a 2D material and modeled by the surface conductivity $\sigma_g$ with the random-phase approximation (RPA) in the local limit, including the intraband and interband transitions\cite{zhang2015towards,xu2015toward}
\begin{equation}\label{eq1}
    \begin{split}
      \sigma_g &=\sigma_{intra}+\sigma_{inter}  \\
               &=\frac{2e^2k_B T}{\pi\hbar^2}\frac{i}{\omega+i\tau^{-1}}\ln[2\cosh(\frac{E_F}{2k_B T})] \\
               &+\frac{e^2}{4\hbar}[\frac{1}{2}+\frac{1}{\pi}\arctan(\frac{\hbar\omega-2E_F}{2k_B T})-\frac{i}{2\pi}\ln\frac{(\hbar\omega+2E_F)^2}{(\hbar\omega-2E_F)^2+4(k_B T)^2}],
    \end{split}
\end{equation}
where $e$ is the charge of an electron, $k_B$ is the Boltzmann constant, $T$ is the operation temperature, $\hbar$ is the reduced Planck's constant, $\omega$ is the angular frequency of the incidence light, $\tau$ is the relaxation time and $E_F$ is the Fermi energy. In the lower THz regime, the interband contributions can be safely neglected due to the Pauli exclusion principle and therefore the surface conductivity can be approximated to a Drude model\cite{zhang2014coherent}
\begin{equation}\label{eq2}
    \sigma_g=\frac{e^2 E_F}{\pi\hbar^2}\frac{i}{\omega+i\tau^{-1}},
\end{equation}
herein the relaxation time $\tau=(\mu E_F)/(e v_F^2)$ depends on the the electron mobility $\mu\approx10000$ cm$^2$/Vs, the Fermi energy $E_F$ and the Fermi velocity $v_F\approx10^6$ m/s. Furthermore, the permitivity of graphene can be obtained by\cite{he2015investigation,vakil2011transformation}
\begin{equation}\label{eq3}
    \varepsilon_g=1+\frac{i \sigma_g}{\varepsilon_0\omega t_g},
\end{equation}
where $\varepsilon_0$ is the permittivity of vacuum.

As predicted in the equation (\ref{eq2}), the surface conductivity of graphene can be tuned via manipulating its Fermi energy. The optical properties of graphene and thus the performances of light trapping in our proposed structure can be dynamically tuned, which can not be realized with metal SPR devices. It is also noted that the most conventional way to manipulate the Fermi energy of graphene, i.e., the backgate configuration would introduce additional electrodes, which makes the fabrication process of the hybrid periodic array much more complicated and is not conducive to practical application. Therefore, in Figure 2 another two feasible methods of tuning $E_F$ are presented: (a) through a uniform external electric field generated by either distant gates or low-frequency radiation\cite{thongrattanasiri2012plasmons,li2014tunable} and (b) through the HNO$_3$ vapor made by external heater\cite{bae2010roll,yan2012tunable}.
\begin{figure}[htbp]\label{fig:2}
\centering
\subfloat[]{
\includegraphics[scale=0.2]{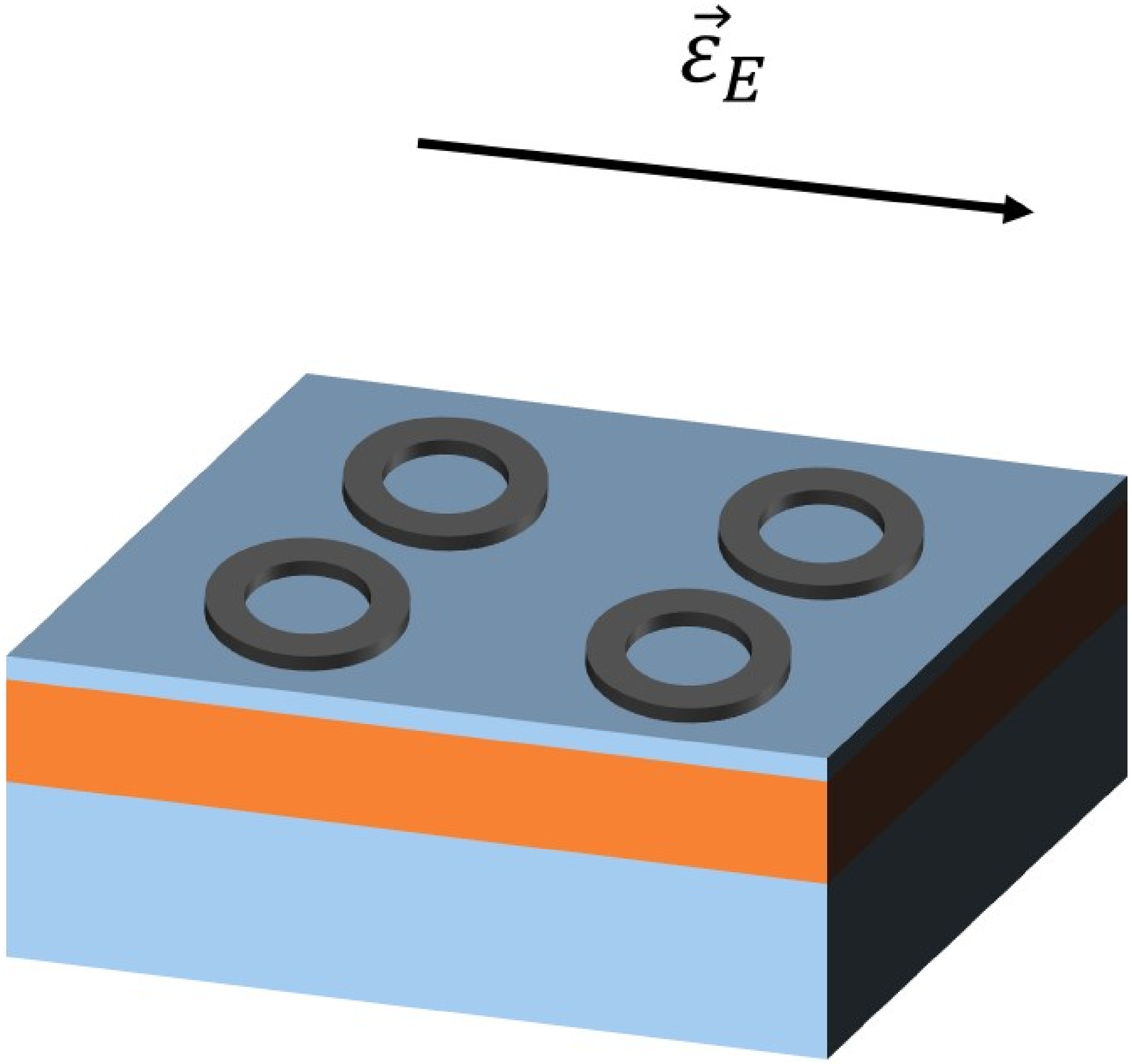}
}
\subfloat[]{
\includegraphics[scale=0.2]{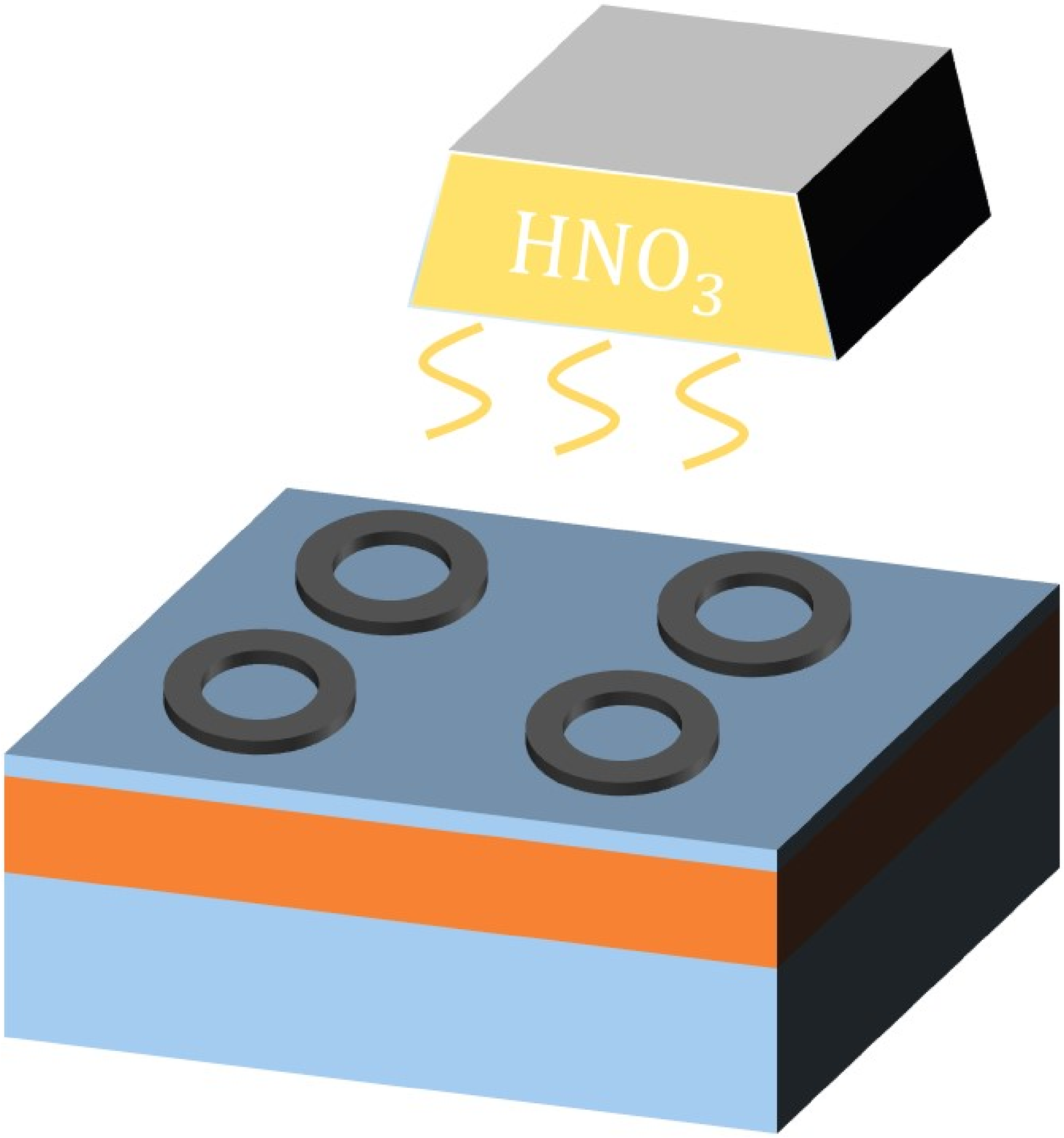}
}
\caption{Two feasible methods of tuning $E_F$: (a) through a uniform external electric field $\vec{\varepsilon}_E$ and (b) through the HNO$_3$ vapor.}
\end{figure}

In the initial setup, the attenuation coefficient of the absorbing layer $\alpha=-0.1$ $\micro\meter$$^{-1}$ and the Fermi energy of graphene $E_F=0.6$ eV are considered, and the influences of them will be analyzed later. The numerical simulations are conducted using the finite-difference time-domain method with the commercial software FDTD Solutions. The anti-symmetric and symmetric boundary conditions are respectively employed in the $x$ and $y$ directions throughout the calculations except when studying the angle polarization tolerance, and perfectly matched layers are utilized in the $z$ direction along the propagation of the incidence plane wave.

\section{Simulation results and discussions}\label{sec3}
For the above structure, a typical resonance response to normal incidence light with transmission suppression and absorption enhancement is displayed at around 19.6 $\micro\meter$ in Figure 3. The total absorption of this structure is $A=30.8\%$ and the absorption in the absorbing layer is $A'=17.4\%$. Now that the absorbing layer is set to 100 nm thick with the attenuation coefficient $\alpha=-0.1$ $\micro\meter$$^{-1}$, which corresponds to a weak absorption of $2\%$ in the impedance matched media, about 8.7 times absorption enhancement in the absorbing layer has been realized at the resonance.
\begin{figure}[htbp]\label{fig:3}
\centering
\includegraphics[scale=0.4]{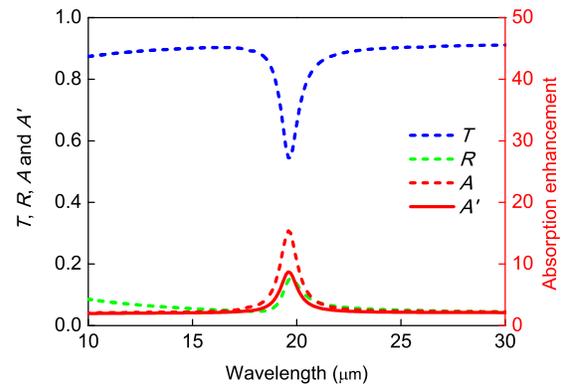}
\caption{The simulated transmission $T$, reflection $R$ and absorption $A$ as well as the absorption in the absorbing layer $A'$ with the attenuation coefficient $\alpha=-0.1$ $\micro\meter$$^{-1}$ and the Fermi energy of graphene $E_F=0.6$ eV. The enhancement factor of absorption in the absorbing layer is also shown compared to that in the impedance matched media.}
\end{figure}

It is not difficult to believe the absorption enhancement should be ascribed to graphene SPR. The localized collective electronic excitations strongly couple to and trap the incidence light and enhance the absorption in the nearby absorbing layer. To better understand and make use of the absorption band, the electric field distributions at the resonance are plotted below.
\begin{figure}[htbp]\label{fig:4}
\centering
\subfloat[]{
\includegraphics[scale=0.3]{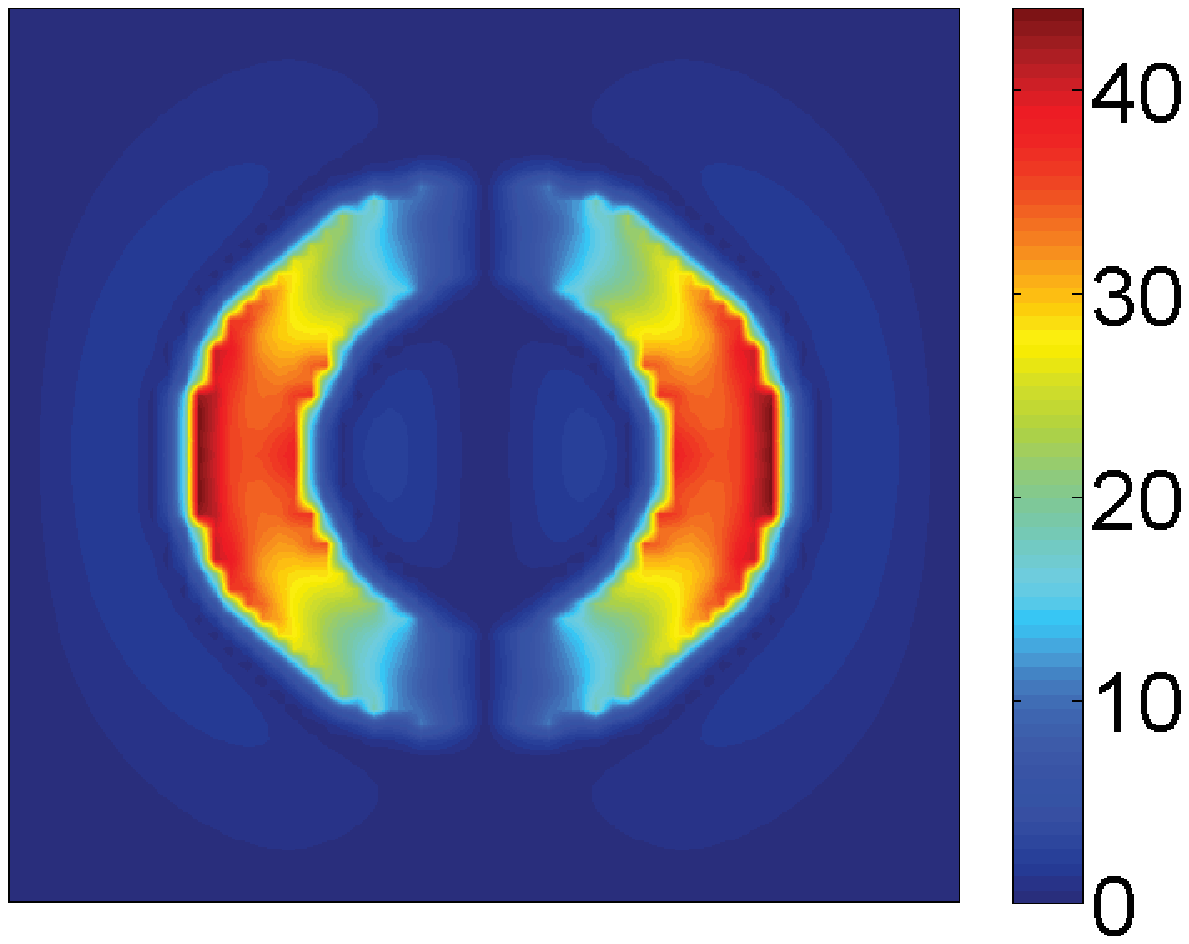}
}
\subfloat[]{
\includegraphics[scale=0.3]{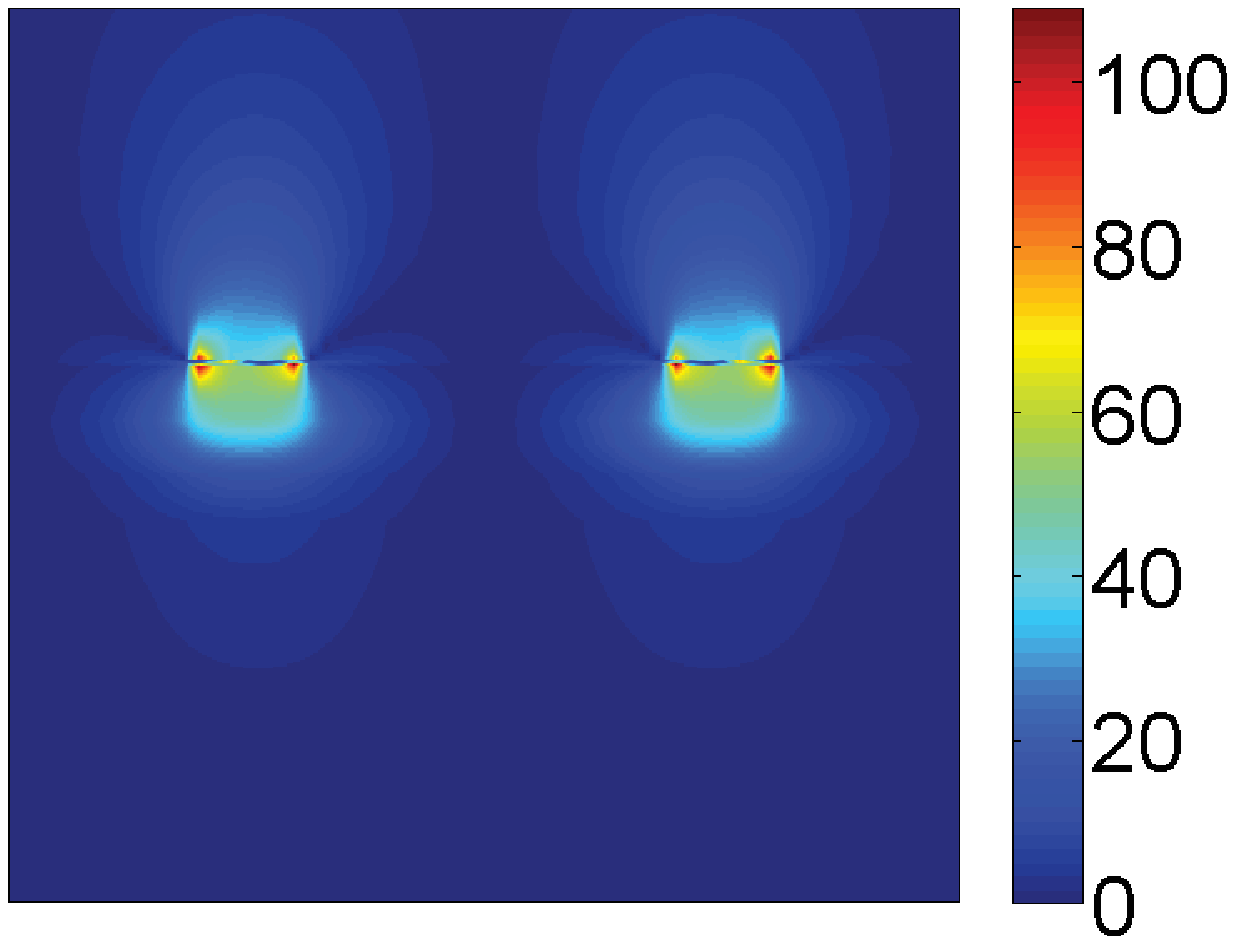}
}
\caption{The simulated electric field distributions (a) in the $x$-$y$ plane ($|E_z|$) and (b) in the $x$-$z$ cross plane ($|E_y|$) at the resonance with the attenuation coefficient $\alpha=-0.1$ $\micro\meter$$^{-1}$ and the Fermi energy of graphene $E_F=0.6$ eV.}
\end{figure}
As shown in Figure 4(a), the $x$-$y$ plane electric field distribution ($|E_z|$) at the resonance shows characteristic behavior of the electric dipole and is symmetric to the $y$-axis due to the $x$-polarization of the incidence light. Meanwhile, the $x$-$z$ cross plane electric field distribution ($|E_y|$) in Figure 4(b) clearly represents that the plasmon-induced strong field confinement extends to the absorbing layer and enhances the light-matter interaction.

Theoretically, it has been known that that the effective wavelength of the above-mentioned dipole resonance is approximately equal to the circumference of the corresponding ring
\begin{equation}\label{eq4}
    \lambda_{eff}=2\pi R,
\end{equation}
and thus the resonance wavelength can be predicted from
\begin{equation}\label{eq5}
    \lambda_{res}=\lambda_{eff} n_{eff}=2\pi R n_{eff},
\end{equation}
where $n_{eff}$ is the effective refractive index of the ring waveguide. For a graphene ribbon waveguide with the same width ($W=30$ nm), $n_{eff}$ at difderent Fermi energies ranging from 0.4 eV - 1.2 eV are plotted in Figure 5 with FDTD Solutions.
\begin{figure}[htbp]\label{fig:5}
\centering
\includegraphics[scale=0.4]{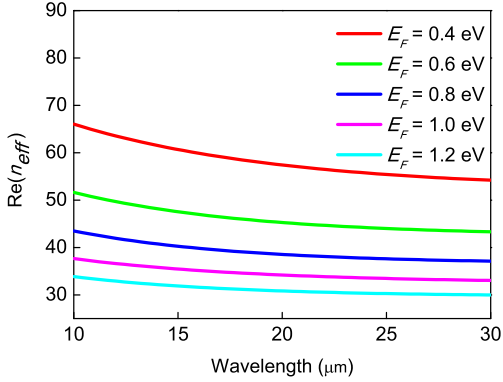}
\caption{The calculated $n_{eff}$ with the attenuation coefficient $\alpha=-0.1$ $\micro\meter$$^{-1}$ and the Fermi energy of graphene $E_F$ ranging from 0.4 to 1.2 eV.}
\end{figure}
It can be seen that $n_{eff}$ is a function of $E_F$, meaning that the resonance wavelength of our proposed structure can be expediently shifted via manipulating the Fermi energy of graphene, which lays the direct foundation for the light trapping and absorption enhancement in the absorbing layer at the selective wavelength.

As the only two lossy media within the hybrid structure, there unavoidably exists a competition of absorption between graphene and the absorbing layer nearby. The attenuation coefficient of the low light-absorbing efficiency material, initially set to $\alpha=-0.1$ $\micro\meter$$^{-1}$, plays a dominant role in the distribution of light absorption. The total absorption $A$ and the absorption in the absorbing layer $A'$ are also simulated with different $\alpha$ in Figure 6. When $\alpha=-0.05$ $\micro\meter$$^{-1}$, corresponding to a poor absorption of $1\%$ in the impedance matched media, the total absorptions increase to $34.6\%$ and the absolute absorption in the absorbing layer is $A'=13.4\%$ with more than one order enhancement at the resonance due to the lower absorption loss of the total structure and the greater quality factor of the resonance. When $\alpha=-0.2$ $\micro\meter$$^{-1}$, corresponding to about $4\%$ absorption in the impedance matched media, the total absorptions decrease to $A=27.0\%$ while the absorption in the absorbing layer is $A'=20.0\%$, in which case about five times enhancement is achieved. It is also noteworthy that absorption band stays at nearly the same resonance wavelength in that the variations in the attenuation coefficient $\alpha$ here are not big enough to bring about visible influences on $n_{eff}$.
\begin{figure}[htbp]\label{fig:6}
\centering
\subfloat[]{
\includegraphics[scale=0.4]{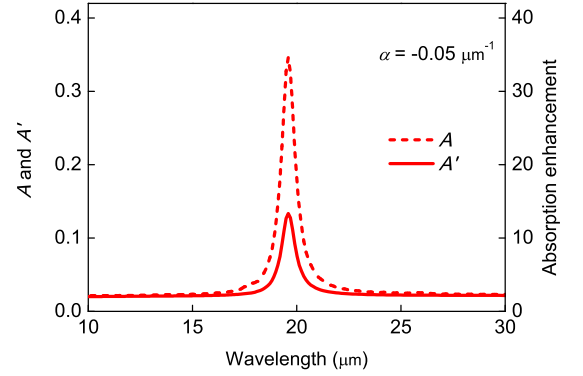}
}\\
\subfloat[]{
\includegraphics[scale=0.4]{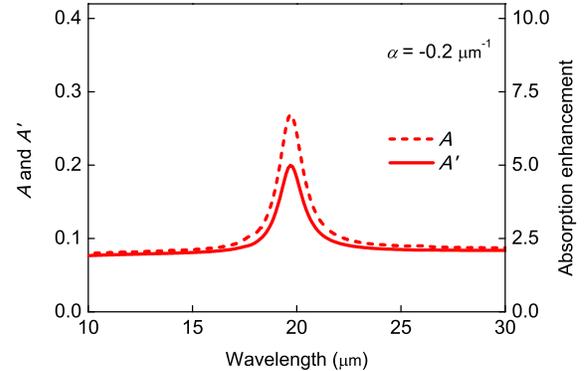}
}
\caption{The simulated absorption $A$ and the absorption in the absorbing layer $A'$ with the attenuation coefficient (a) $\alpha=-0.05$ $\micro\meter$$^{-1}$ and (b) $\alpha=-0.2$ $\micro\meter$$^{-1}$ and the Fermi energy of graphene $E_F=0.6$ eV. The enhancement factor of absorption in the absorbing layer is also shown compared to that in the impedance matched media.}
\end{figure}

To analyze the tunable property of the light trapping, the absorption in the absorbing layer with various Fermi energy of graphene has been simulated in Figure 7. When the Fermi energy starting at 0.4 eV, the resonance wavelength is 24.0 $\micro\meter$ and the absorption is $11.3\%$. As $E_F$ increases to 0.8 eV, the resonance shifts to 17.1 $\micro\meter$ and the absorption goes up to 22.1\% with more than one order absorption enhancement. Finally while $E_F$ comes to 1.2 eV, the resonance absorption reaches as high as $27.7\%$ at 13.8 $\micro\meter$. The physical mechanism lies in that the conductivity of graphene increases with $E_F$ and thus graphene SPR becomes less lossy, which leads to more absorption in the absorbing layer.
\begin{figure}[htbp]\label{fig:7}
\centering
\includegraphics[scale=0.4]{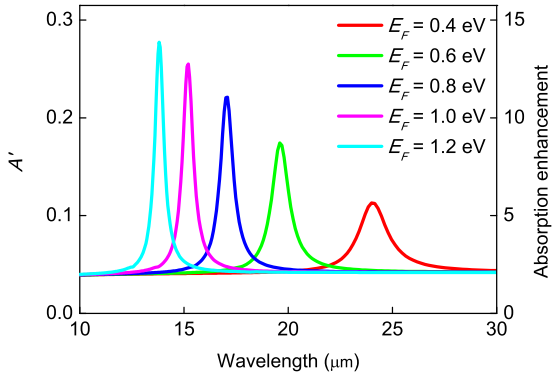}
\caption{The simulated absorption in the absorbing layer $A'$ with the attenuation coefficient $\alpha=-0.1$ $\micro\meter$$^{-1}$ and the Fermi energy of graphene $E_F$ ranging from 0.4 to 1.2 eV. The enhancement factor of absorption in the absorbing layer is also shown compared to that in the impedance matched media.}
\end{figure}
The resonance wavelengths in theoretical analysis (red line) and simulation (green line) are plotted in Figure 8. One can see the theoretical resonance wavelengths at different Fermi energies, calculated with the equation (\ref{eq5}), follow the fairly similar trend as the simulation results, though minute differences exist. This could be because the $n_{eff}$ was calculated in the graphene ribbon waveguide, which has some geometric differences with the ring waveguide in our proposed structure, and the deviation of $n_{eff}$ should lead to the deviation of the resonance wavelength. Considering this, the agreement between theory and simulation is reasonable and could be employed in predicting the resonance wavelength of our proposed device for light trapping and absorption enhancement. As a consequence, photoresponse amplification with high efficiency and tunable spectral selectivity can be achieved with graphene SPR for highly accurate photodetection, which can not be realized with metal SPR.
\begin{figure}[htbp]\label{fig:8}
\centering
\includegraphics[scale=0.4]{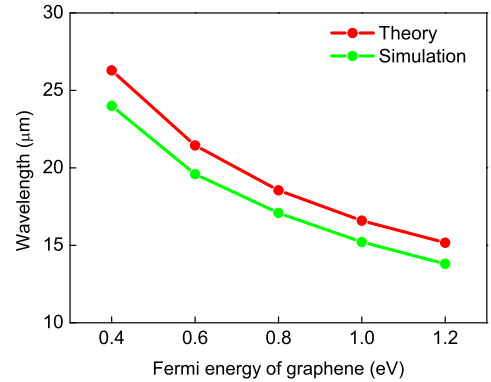}
\caption{The resonance wavelengths in theoretical analysis and simulation with the attenuation coefficient $\alpha=-0.1$ $\micro\meter$$^{-1}$ and the Fermi energy of graphene $E_F$ ranging from 0.4 to 1.2 eV.}
\end{figure}

Considering the high symmetry of our proposed structure, the incidence angle and polarization dependence of graphene ring is expected to be weak. The light absorption properties at various angle for both TE and TM configurations have been investigated. As illustrated in Figure 9(a) and (b), the absorption in the absorbing layer keeps nearly the same for incidence angle up to 45 degrees. On the other hand, the resonance wavelength is only determined by the radius and the Fermi energy of the graphene ring, indicated by the equation (\ref{eq5}), and does not depend on the incidence angle and polarization, which is also reflected in the figures. Consequently, the good operation angle and polarization tolerance of graphene ring will benefit the practical applications in light trapping and absorption enhancement.
\begin{figure}[htbp]\label{fig:9}
\centering
\subfloat[]{
\includegraphics[scale=0.3]{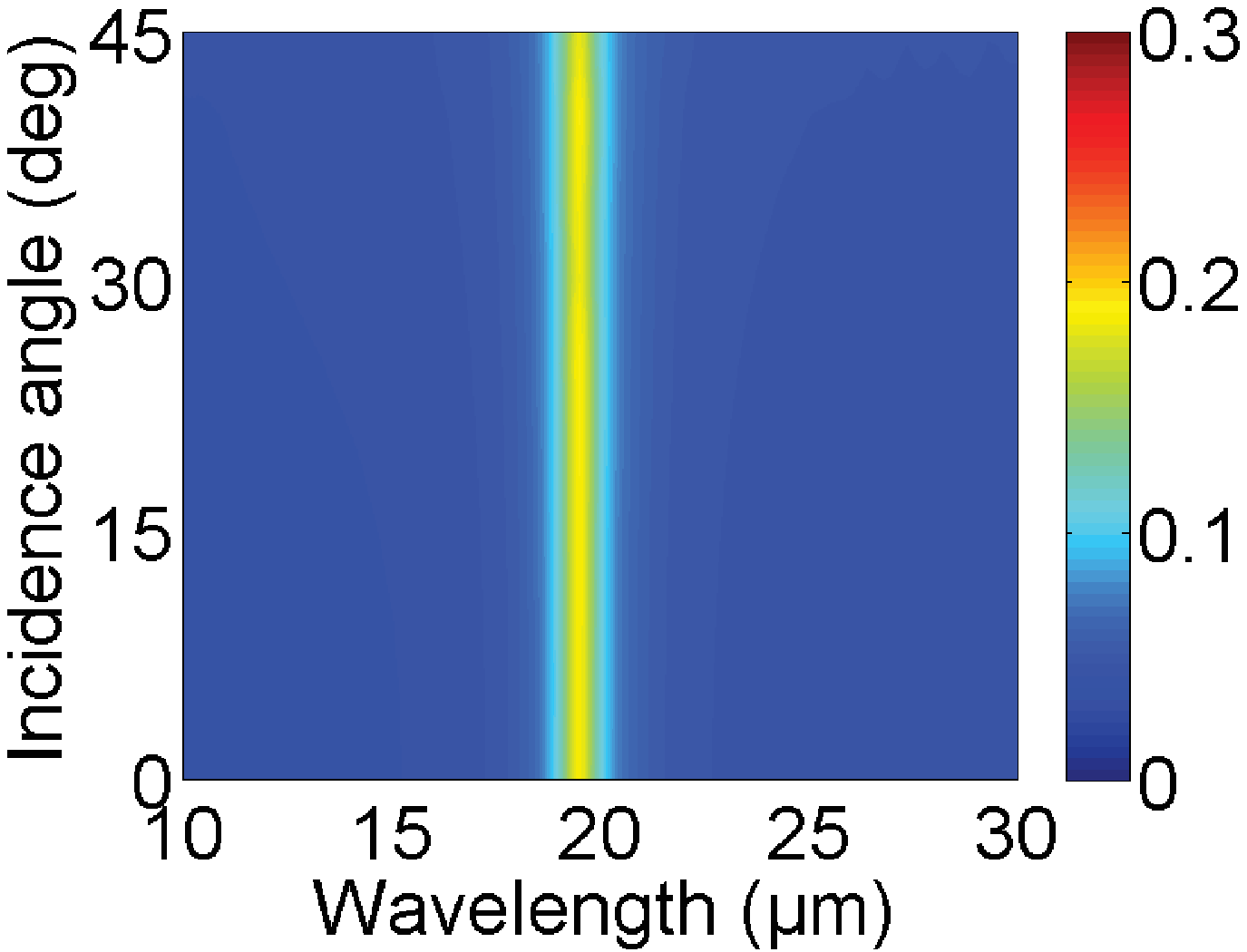}
}
\subfloat[]{
\includegraphics[scale=0.3]{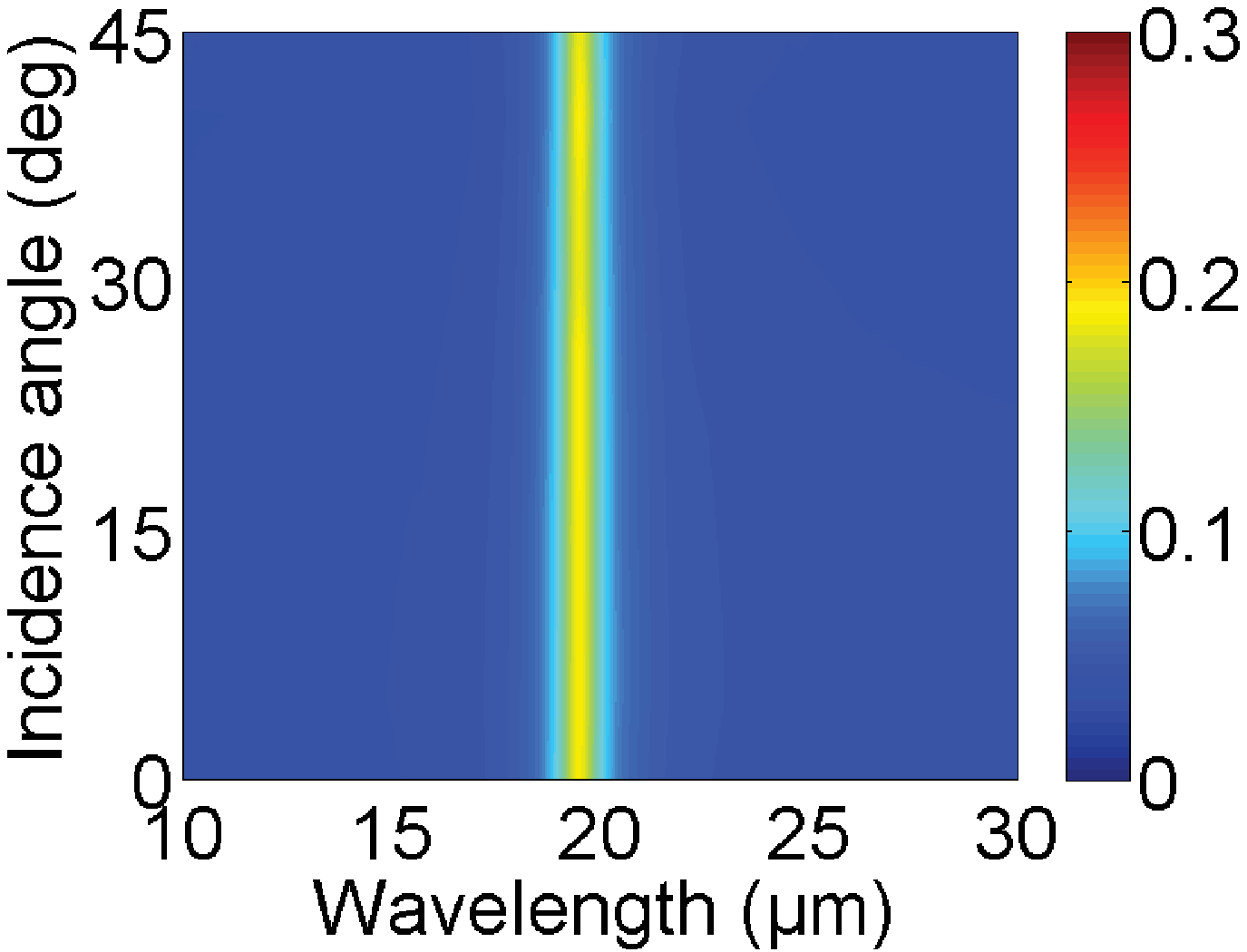}
}
\caption{The simulated angular dispersions of the absorption in the absorbing layer with the attenuation coefficient $\alpha=-0.1$ $\micro\meter$$^{-1}$ and the Fermi energy of graphene $E_F=0.6$ eV for (a) TE and (b) TM configurations.}
\end{figure}

In the most recent years, 2D materials have emerged as promising building blocks in the state of the art optoelectronic devices among a very wide range of electromagnetic spectrum\cite{xia2014two}. Compared with the traditional photonic materials, 2D materials exhibit a lot of exceptional properties. The quantum confinement in the direction perpendicular to the 2D plane leads to many novel optical and electronic properties\cite{mak2010atomically,novoselov2005two,zhang2005experimental,splendiani2010emerging}, and the naturally passivated surfaces make them easy to integrate with photonic structures such as waveguides\cite{liu2011graphene,gan2013chip} and cavities\cite{gan2012strong,gan2013controlling}. In addition to graphene, a variety of other 2D materials such as transition metal dichalcogenides have been employed in photodetection and other optoelectronic applications \cite{lopez2013ultrasensitive,wang2015hot}. Though possessing very high quantum efficiency for light-matter interaction, the absorption of these 2D materials is insufficient in practical devices due to the innate thinness. Therefore, integration with SPR offers an effective solution for the absorption enhancement and further applications in photodetection. As shown in Figure 10, the 100 nm thick absorbing layer is replaced with 2D material, which is modelled as a 1nm thick conductive film with an optical conductance $G_0=2.65\times10^{-5}$ $\Omega^{-1}$, corresponding to an absorption of about $1\%$ for a free standing film.
\begin{figure}[htbp]\label{fig:10}
\centering
\includegraphics[scale=0.4]{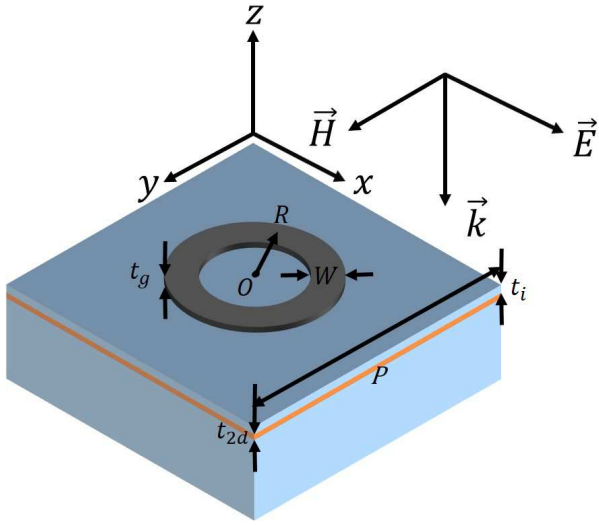}
\caption{The schematic geometry of the unit cell composed of a graphene ring on the top of the absorbing layer separated by an insulating layer and thin film semiconductor is replaced with 2D material.}
\end{figure}
Similar to that in the thin film semiconductor, the resonance response is also found here. Figure 11 plots the absorption enhancement in the 2D absorbing layer while Fermi energy of graphene varies from 0.4 eV to 1.2 eV in step of 0.2 eV. When $E_F$ starting at 0.4 eV, the resonance wavelength locates at 20.5 $\micro\meter$, and the absorption is 7.9\% with an enhancement factor of 7.9. When $E_F$ increases to 0.6 eV, the resonance blue shifts to 16.6 $\micro\meter$ and the absorption goes up to 14.3\% with more than one order absorption enhancement. And while $E_F$ comes to 1.2 eV, the absorption reaches as high as $25.2\%$ at 11.7 $\micro\meter$ and the enhancement factor over 25 is obtained at the resonance.
\begin{figure}[htbp]\label{fig:11}
\centering
\includegraphics[scale=0.4]{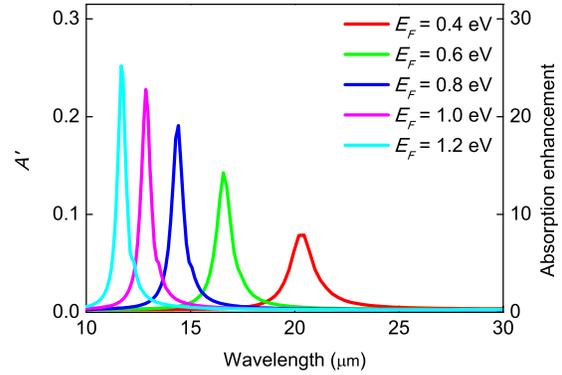}
\caption{The simulated absorption in the 2D material $A'$ with an optical conductance $G_0=2.65\times10^{-5}$ $\Omega^{-1}$ and the Fermi energy of graphene $E_F$ ranging from 0.4 to 1.2 eV. The enhancement factor of absorption in the absorbing layer is also shown compared to that in the impedance matched media.}
\end{figure}

In the above simulations, the geometric parameters of the hybrid structure are fixed and the tunability is mainly demonstrated via manipulating the Fermi energy of graphene. In fact, the light trapping and absorption enhancement could also be tuned with the geometric variations of the structure (See ESI Figure S1). It is also noted that the total thickness of the hybrid structure is much smaller than the wavelength of the incidence light, in which case the universal maximum absorption can be evaluated with regarding the whole structure as a thin film separating two different media (air with the refractive index of $n_{air}=1$ and the substrate with $n_{sub}=1.4$), i.e., $A_{max}=41.7\%(=1/(1+1.4))$\cite{thongrattanasiri2012complete}. Perfect light absorption in the hybrid structure and further absorption enhancement in the absorbing layer are obtained with a gold mirror (See ESI Figure S2). Moreover, another major advantage of our proposed structure lies in that it is straightforward to include more rings into the unit cell to form a compact structure with multiple resonances. For example, the light trapping and absorption enhancement with a pair of graphene concentric rings are studied. Figure 12 depicts the schematic geometry of the structure composed of a pair of graphene concentric rings on the top of the thin film semiconductor separated by a insulating layer. The inner graphene ring remains as $R_1=R=75$ nm and the outer one is added as $R_2=125$ nm, and the attenuation coefficient and other geometric parameters keep the same.
\begin{figure}[htbp]\label{fig:12}
\centering
\includegraphics[scale=0.4]{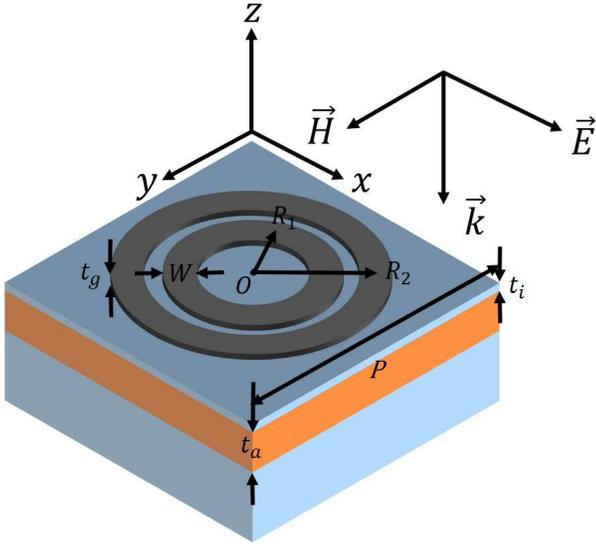}
\caption{The schematic geometry of the unit cell composed of a pair of graphene concentric rings on the top of the absorbing layer separated by an insulating layer.}
\end{figure}

Figure 13 illustrates the absorption in the absorbing layer at different Fermi energies. The results show that another absorption band, as expected, appears at longer wavelength, and when the Fermi energy gets to 1.0 eV, the absorption goes up to $21.7\%$ at 26.9 $\micro\meter$, which implies more than one order absorption enhancement is achieved simultaneously at both resonances. More interesting, for each absorption band at shorter wavelength, the absorption keeps nearly the same strength and the resonance wavelength stays nearly the same position as with single graphene ring, which suggest each resonance could be approximated to individual dipole response of the corresponding graphene ring.
\begin{figure}[htbp]\label{fig:13}
\centering
\includegraphics[scale=0.4]{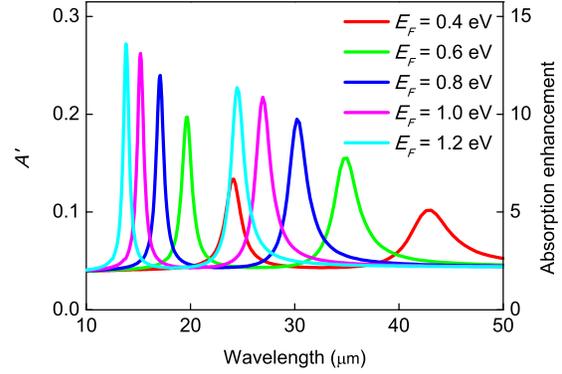}
\caption{The simulated absorption in the absorbing layer $A'$ with the attenuation coefficient $\alpha=-0.1$ $\micro\meter$$^{-1}$ and the Fermi energy of graphene $E_F$ ranging from 0.4 to 1.2 eV. The enhancement factor of absorption in the absorbing layer is also shown compared to that in the impedance matched media.}
\end{figure}
To confirm this, the $x$-$y$ plane electric field distribution ($|E_z|$) and the $x$-$z$ cross plane electric field distribution ($|E_y|$) with the attenuation coefficient $\alpha=-0.1$ $\micro\meter$$^{-1}$ and the Fermi energy of graphene $E_F=0.6$ eV at the resonances are presented in Figure 14. One can tell the electric field mainly concentrates in either inner or outer ring and the mutual couplings between them are rather weak. Therefore, the design principle here could be safely set as a template to achieve tunable multiband plasmonic absorption enhancement in the below light-absorbing materials, and consequently meets the urgent need for the simultaneous multi-color photodetection with high efficiency and tunable spectral selectivity.
\begin{figure}[htbp]\label{fig:14}
\centering
\subfloat[]{
\includegraphics[scale=0.3]{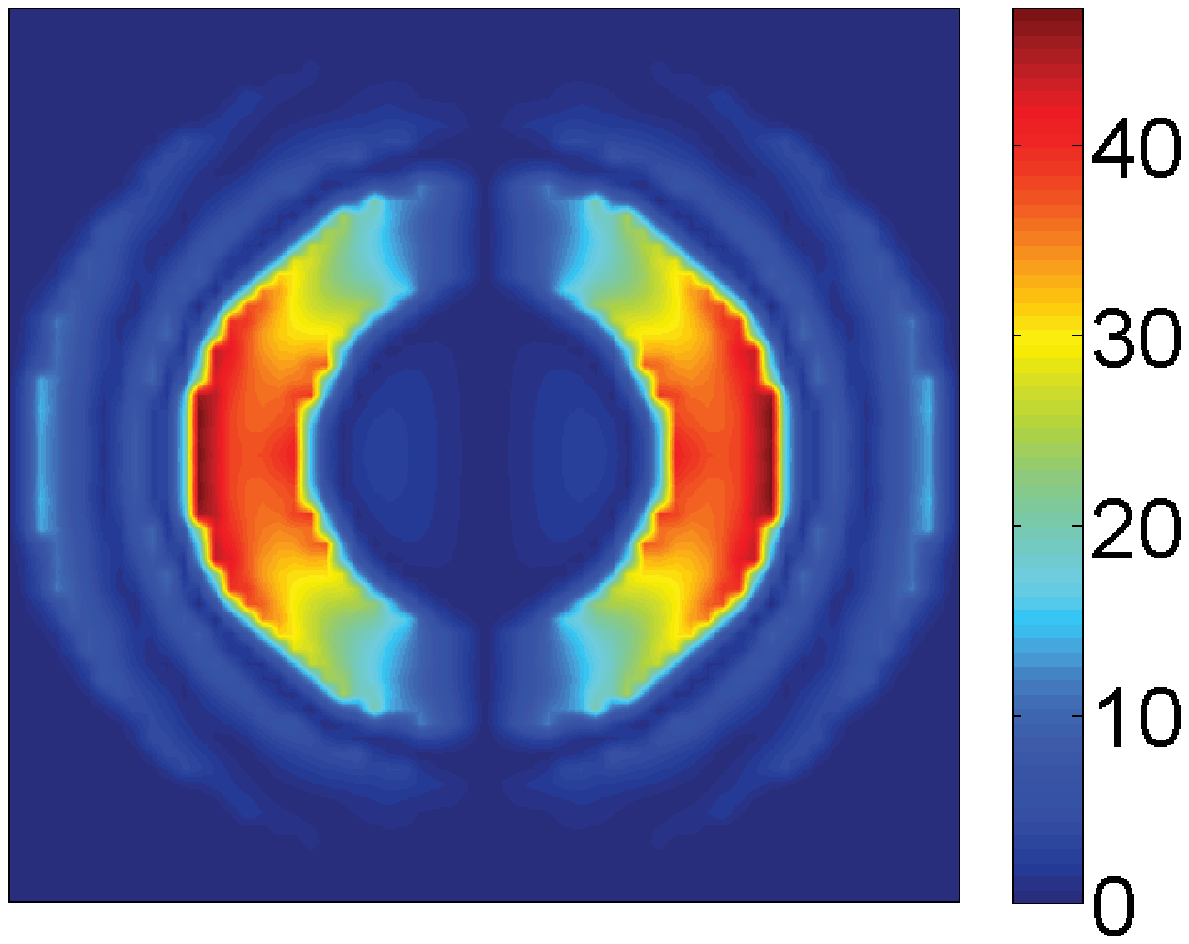}
}
\subfloat[]{
\includegraphics[scale=0.3]{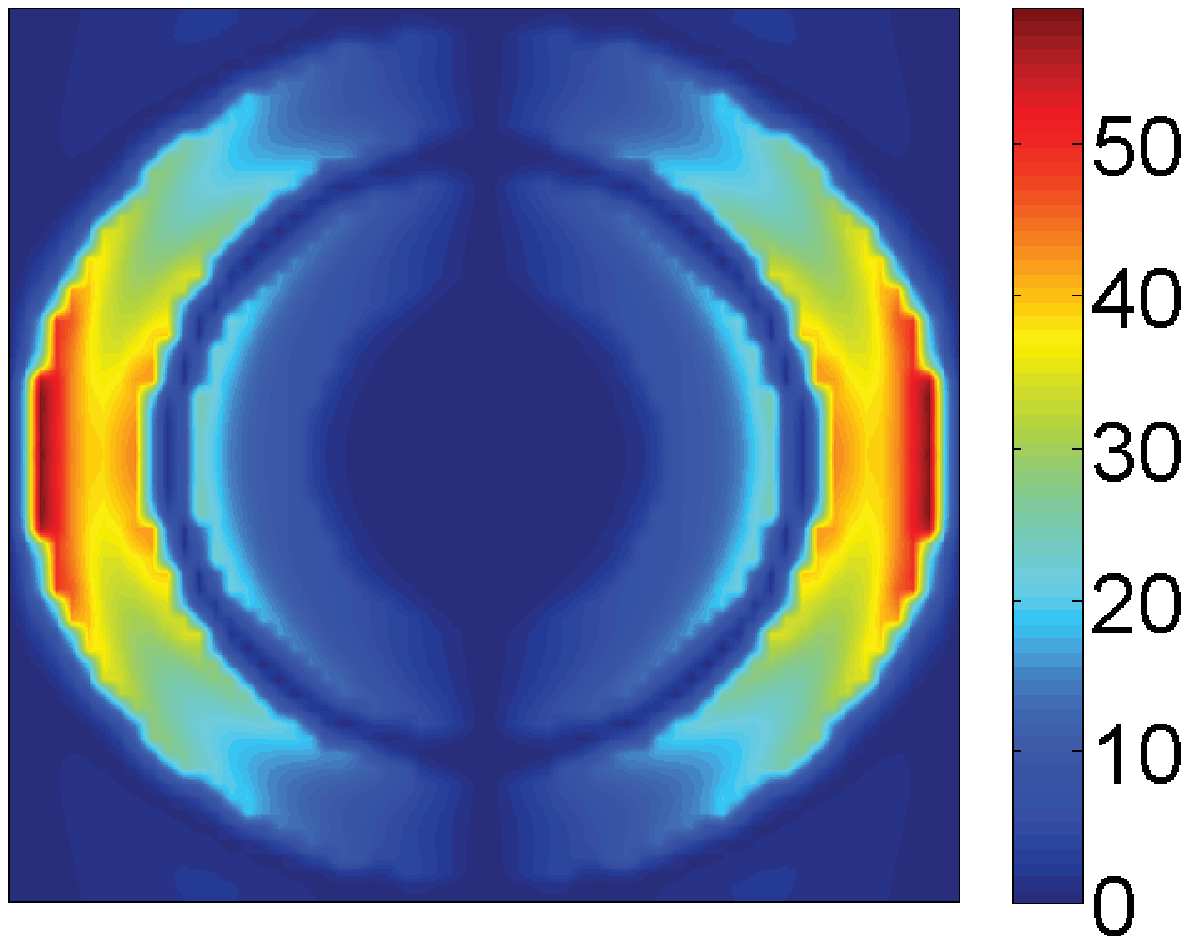}
}\\
\subfloat[]{
\includegraphics[scale=0.3]{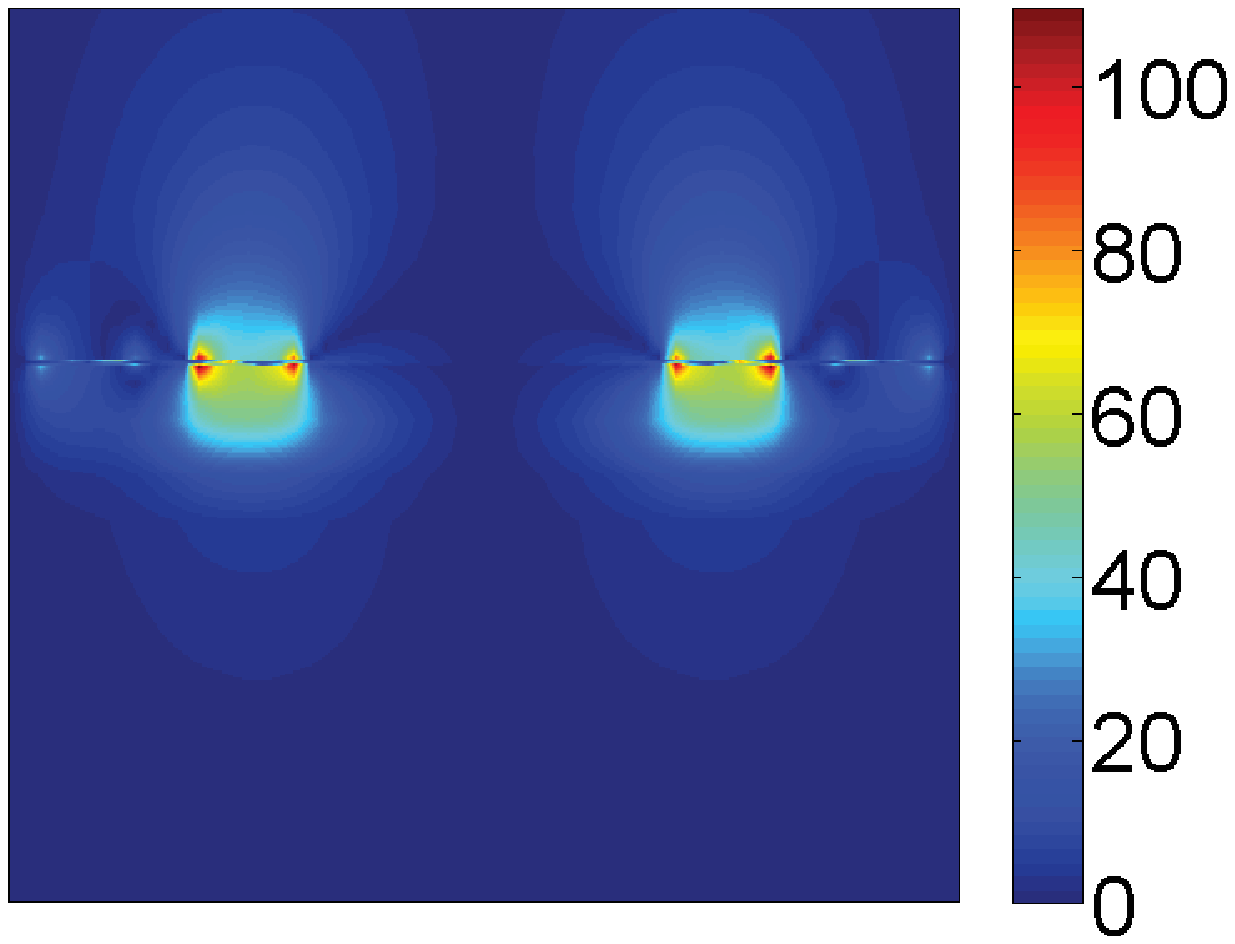}
}
\subfloat[]{
\includegraphics[scale=0.3]{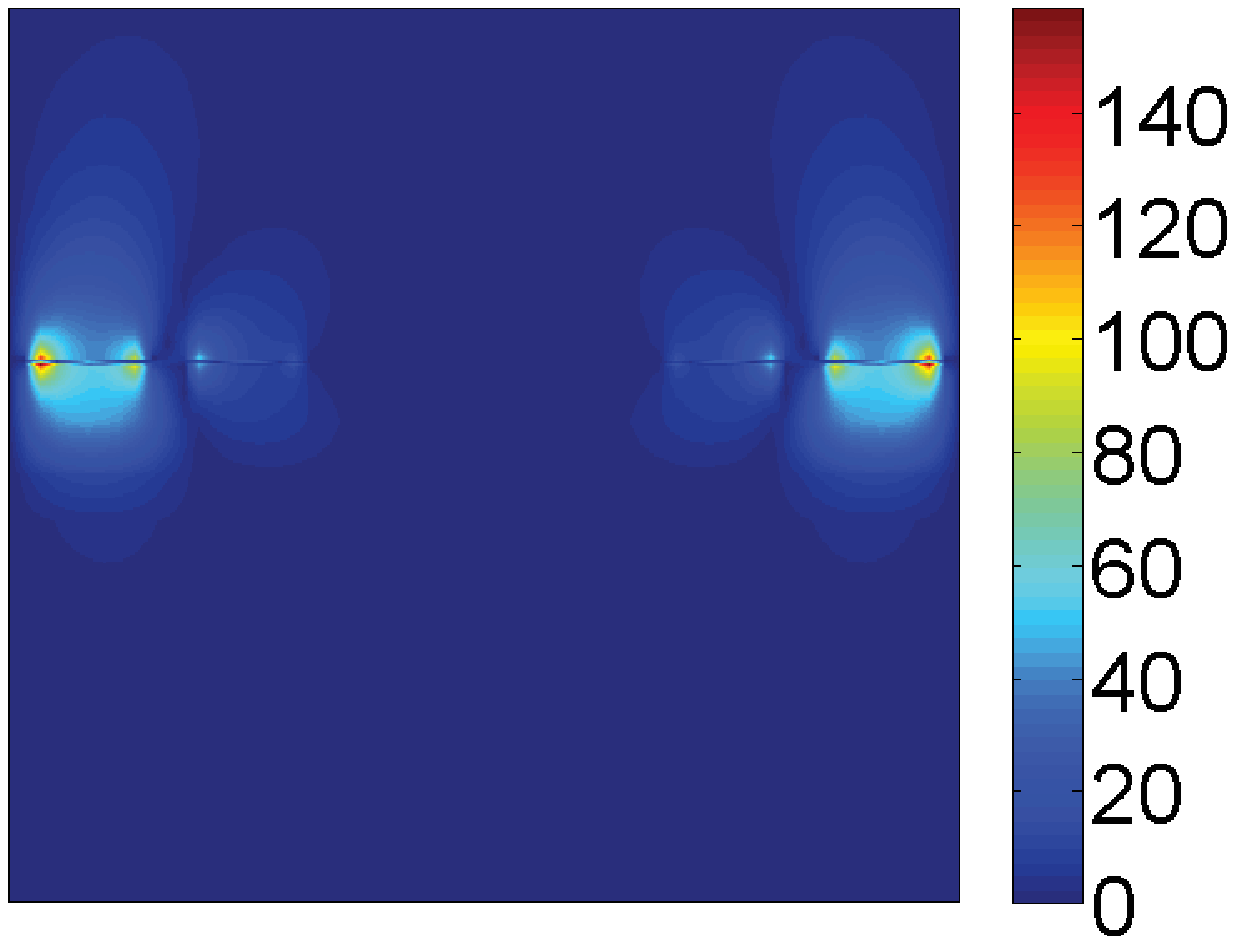}
}
\caption{The simulated electric field distributions (a)-(b) in the $x$-$y$ plane ($|E_z|$) and (c)-(d) in the $x$-$z$ cross plane ($|E_y|$) at the resonances with the attenuation coefficient $\alpha=-0.1$ $\micro\meter$$^{-1}$ and the Fermi energy of graphene $E_F=0.6$ eV.}
\end{figure}

\section{Conclusions}\label{sec4}
To conclude, tunable light trapping and absorption enhancement have been numerically investigated in a hybrid periodic array composed of a graphene ring on the top of the absorbing layer separated by an insulating layer. It is found that the excitations of localized collective electrons in graphene strongly couple to and trap the incidence light in the near-field, and significantly enhance the absorption in the light-absorbing materials (thin film semiconductor or 2D material) to more than one order at the resonance. Compared with the traditional noble metal plasmonic devices, the absorption enhancement here can be dynamically tuned via manipulating the Fermi energy of graphene over such a large range, thus enables tunable photodetection and other optoelectronic applications in the mid-infrared and THz regime. As an extension, our proposed structure leaves room for introducing more graphene concentric rings to the unit cell to achieve multi-band absorption enhancement, therefore lays the foundation for simultaneous photodetection at multiple wavelengths with high efficiency and tunable spectral selectivity.

\section*{Acknowledgments}
The author Shuyuan Xiao (SYXIAO) expresses his deepest gratitude to his Ph.D. advisor Tao Wang for providing guidance during this project. SYXIAO would also like to thank Prof. Jianfa Zhang (National University of Defense Technology) for his guidance to the modeling of light-absorbing materials and Dr. Qi Lin (Hunan Univerisity) for beneficial discussion on graphene optical properties. This work is supported by the National Natural Science Foundation of China (Grant No. 61376055 and 61006045), and the Fundamental Research Funds for the Central Universities (HUST: 2016YXMS024).

\balance

\footnotesize{
\bibliography{rsc} 
\bibliographystyle{rsc} 

}

\end{document}